\documentclass[nodraftcls,conference]{IEEEtran}

\usepackage{amsmath,graphicx}
\usepackage{amssymb}
\usepackage{algpseudocode}
\usepackage{cite}
\usepackage{amsfonts}
\usepackage{graphicx,subfigure}
\usepackage{bm}

\usepackage{fancyhdr}  %
\usepackage{cases}
\usepackage{extarrows}
\usepackage{algorithm}
\usepackage{multirow,tabularx}
\usepackage{mathtools}
\usepackage{xcolor}
\usepackage[english]{babel}

\PassOptionsToPackage{bookmarks=false}{hyperref}
\PassOptionsToPackage{hyperindex=false}{hyperref}
\PassOptionsToPackage{link=false}{hyperref}
\PassOptionsToPackage{linktoc=false}{hyperref}

\IEEEoverridecommandlockouts   

\begin{document}

\title{Indoor Signal Focusing with Deep Learning Designed Reconfigurable Intelligent Surfaces}
\author{
\IEEEauthorblockN{Chongwen~Huang$^1$, George~C.~Alexandropoulos$^2$, Chau Yuen$^1$, and M\'{e}rouane Debbah$^{3,4}$
\thanks{The work of C. Yuen was supported by the MIT-SUTD International design center and NSFC 61750110529 Grant, and that of C. Huang by the PHC Merlion PhD program. The work of M. Debbah was supported by H2020 MSCA IF BESMART (Grant 749336) and H2020-ERC PoC-CacheMire (727682).}
 }
\IEEEauthorblockA{
$^1$Singapore University of Technology and Design, 487372 Singapore\\
$^2$Department of Informatics and Telecommunications, National and Kapodistrian University of Athens, Greece\\
$^3$CentraleSup\'elec, Universit\'e  Paris-Saclay, 91192 Gif-sur-Yvette, France\\
$^4$Mathematical and Algorithmic Sciences Lab, Huawei Technologies France SASU,92100 Boulogne-Billancourt,France\\
emails: chongwen\_huang@mymail.sutd.edu.sg, alexandg@di.uoa.gr, yuenchau@sutd.edu.sg, merouane.debbah@huawei.com
}}

\maketitle


\begin{abstract}
Reconfigurable Intelligent Surfaces (RISs) comprised of tunable unit elements have been recently considered in indoor communication environments for focusing signal reflections to intended user locations. However, the current proofs of concept require complex operations for the RIS configuration, which are mainly realized via wired control connections. In this paper, we present a deep learning method for efficient online wireless configuration of RISs when deployed in indoor communication environments. According to the proposed method, a database of coordinate fingerprints is implemented during an offline training phase. This fingerprinting database is used to train the weights and bias of a properly designed Deep Neural Network (DNN), whose role is to unveil the mapping between the measured coordinate information at a user location and the configuration of the RIS's unit cells that maximizes this user's received signal strength. During the online phase of the presented method, the trained DNN is fed with the measured position information at the target user to output the optimal phase configurations of the RIS for signal power focusing on this intended location. Our realistic simulation results using ray tracing on a three dimensional indoor environment demonstrate that the proposed DNN-based configuration method exhibits its merits for all considered cases, and effectively increases the achievable throughput at the target user location.
\end{abstract}



\section{Introduction}

Mobile internet service has spurred exponential growth in cellular network usage, such as video sharing, virtual/augmented reality, and movie downloading. The explosion of data traffic volume has been further fueled by smartphones and various other portable devices \cite{fetocell2013,Heter2011}. However, a common complaint with current cellular systems is their imperfect coverage, especially indoors. Due to the high penetration loss from walls, the signal strength received from an outdoor Base Station (BS) inside a building may be too low to attain acceptable performance \cite{Heter2011,husha_LIS1}. Since BSs cannot increase transmission power unlimitedly to enhance received signals in outage areas, a better solution may be to deploy denser transmission points to cover these locations. This refers to the deployment of additional BSs or multiple relays, which is however economically inefficient due to the tedious location finding process and the high backhaul, power consumption, as well as implementation cost \cite{fetocell2013,husha_LIS1,LIS_ICASSP2018,qingqing2019,LIS_globecom2018}.

A recent emerging hardware technology named Reconfigurable Intelligent Surface (RIS) offers a promising solution for
indoor communications \cite{husha_LIS1,LIS_globecom2018,qingqing2019}. A RIS is a thin surface equipped with integrated electronic circuits that can be programmed to alter an incoming electromagnetic field in a customizable way. It consists of a single- or few-layer stack of planar structures that can be readily fabricated using lithography and nano-printing methods. The RIS units are implemented by reflecting elements that employ varactor diodes or other micro-electrical-mechanical systems, and whose resonant frequency is electronically controlled \cite{LIS_globecom2018,qingqing2019}. The natural vision with this technology is the realization of intelligent environments where RISs effectively act as transmitting and receiving structures enabling highly accurate focusing of the transmitted energy to intended user locations. This feature will lead to unprecedented data rates at targeted locations in the 3-Dimensional (3D) space, while avoiding interference at nearby locations.

To strengthen signal focusing in indoor environments, there has been a variety of wireless technologies proposed in the past few decades. One of the first approaches detailed in Hansen's book \cite{hansen1964} utilizes a dedicated antenna array for improving beamforming to intended user positions. In \cite{fetocell2013}, WiMAX femtocells were proposed for increasing indoor coverage, while \cite{Heter2011} presented a heterogeneous network consisting of a relay station, a WiFi Access Point (AP) and a macrocell BS which can collectively provide significant area capacity gain and indoor coverage improvement. In addition, the fresnel zone plate lens were utilized in \cite{Kishk2009} as focused antennas to enhance signal focusing. Whether multiple transmit antennas or fresnel zone plate lens deployed for signal focusing, most of the existing approaches employ traditional beamforming technology that require dedicated hardware and signal processing algorithms. Furthermore, a variety of machine learning algorithms have been applied for achieving highly accurate signal focusing. Among them are the K-nearest neighbors, support vector machines, and Bayesian learning \cite{Loc_song2018}. Very recently, deep learning methods have demonstrated significant improvements in various applications (e.g., \cite{Alex_2018,ML_ICC2019}). These methods are capable of outperforming human-level object detection in some tasks, achieving the state-of-the-art results in machine translation and speech processing, as well as improving outdoor localization in massive antenna systems \cite{Alex_2018}. However, indoor signal focusing based on deep learning configured RISs for avoiding their complex wireless tuning has not yet reported in the literature.

In this paper, we propose a deep learning method for efficient online configuration of RISs aiming to improve the transmitted signal focusing to intended receiver positions in indoor communication environments. A Deep Neural Network (DNN) is designed to learn the mapping between the measured position information at a user location and the optimal configuration of the RIS's unit cells that maximizes this user's received signal strength. We construct a fingerprinting database during an offline phase to train the designed DNN, which consists of the measured position information at reference positions and their corresponding optimal RIS configurations. In the online phase, the trained DNN is utilized to predict the optimal tuning of the RIS's elements given the measured position at the targeted user location. It is shown that the proposed DNN-based approach enables efficient online wireless RIS configuration offering improved signal focusing, and thus increased achievable rates, in desired points in the 3D space with lower estimated errors.


\section{System Model And Proposed Methodology}


\subsection{System Model}

\begin{figure}
  \begin{center}
   \vspace{-0mm}
  \includegraphics[width=85mm]{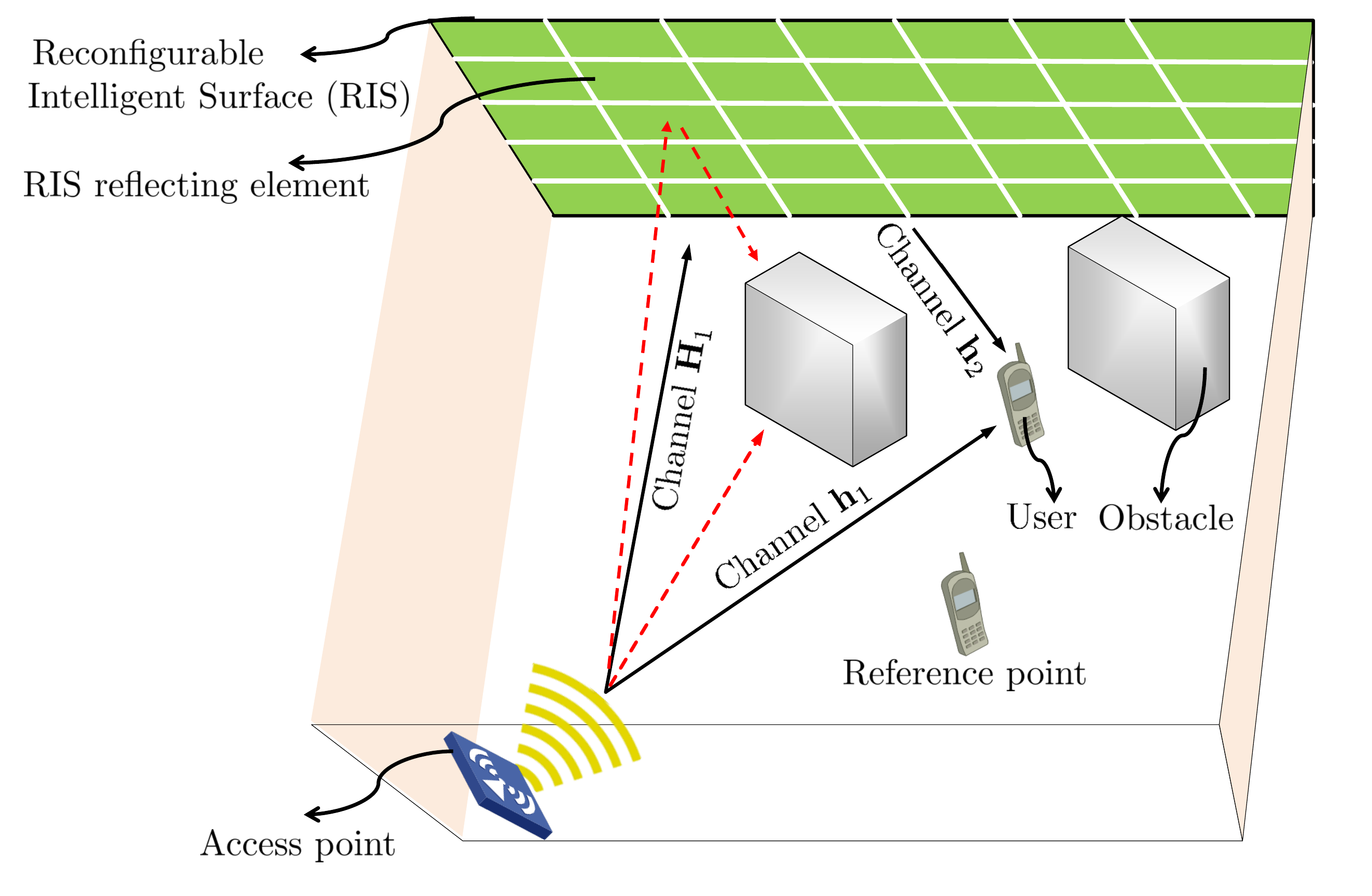}
  \caption{The considered RIS-assisted indoor communication system.}
  \label{fig:lisindoor} \vspace{-8mm}
  \end{center}
\end{figure}
Consider a 3D indoor environment, as depicted in Fig$.$~\ref{fig:lisindoor}, including one AP equipped with $M$ antenna elements and a RIS comprising of $N$ reflecting elements. Different from our previous works \cite{LIS_globecom2018,LIS_twc2018}, we assume that the RIS is also equipped with few active elements that enable estimation of the channels where the RIS is involved. This estimation can take place indirectly via configuration searching or by using sophisticated signal processing algorithms (e.g., via channel matrix completion \cite{George2018}); this challenging task is left for future work. In Fig$.$~\ref{fig:lisindoor}, we illustrate one Reference Point (RP) and two obstacles as an example, but we actually deploy many RPs to collect the fingerprints on a database. All reference points and user locations are assumed on the room's floor. In the figure there exists a target user equipped with a wireless device capable of performing estimation of the channel between itself and AP as well as with RIS using classical channel estimation techniques. We assume 3D signal propagation from AP and the RIS reflections, which we will focus by appropriate RIS configuration on intended user's position. For simplicity, we also assume that ray tracings will be absorbed when they touch the ceiling, floor and walls of the indoor environment except the RIS. The absorbed signal rays are denoted by dotted red lines in Fig$.$~\ref{fig:lisindoor}. The complex-valued baseband (discrete time) received signal at the intended user can be mathematically expressed as
\begin{equation}\label{model01}
  z \triangleq \mathbf{h}_{2}\mathbf{\Phi}\mathbf{H}_{1}\mathbf{s}+\mathbf{h}_1\mathbf{s}+n,
\end{equation}
where $\mathbf{h}_{2,k}\in\mathbb{C}^{1\times N}$ denotes the channel vector between RIS and the user, $\mathbf{H}_{1} \in\mathbb{C}^{N\times M}$ represents the channel matrix between AP and RIS, $\mathbf{h}_1\in\mathbb{C}^{1\times M}$ is the direct channel matrix between AP and the user, and $\mathbf{\Phi}\triangleq\mathrm{diag}[\phi_1,\phi_2,\ldots,\phi_N]$ denotes a diagonal matrix including the phase shifts that are effectively applied by the RIS reflecting elements. In addition, $\mathbf{s}\triangleq\sqrt{\frac{p}{M}}\mathbf{1}_Ms\in\mathbb{C}^{M\times1}$ represents the AP's transmitted signal, where $\mathbf{1}_M$ is the column vector with $M$ ones, $p$ denotes the transmit power, and $s$ is the unit power information symbol chosen from a discrete constellation set. Finally, $n\sim\mathcal{CN}(0,\sigma^{2})$ in \eqref{model01} models the thermal noise at the user's receiver. For the wireless channels we assume that the large-scale fading (shadowing) is constant with respect to frequency, and that the small-scale fading follows a complex Gaussian distribution. It easily follows from \eqref{model01} that the Signal to Noise Ratio (SNR) at the intended user can be expressed as
\begin{align}\label{model_06}
  \gamma \triangleq \frac{p}{\sigma^2}\left|\mathbf{h}_{2}\mathbf{\Phi} \mathbf{H}_{1}+\mathbf{h}_{1}\right|^2,
\end{align}
hence, the achievable throughput performance of the considered RIS-assisted communication is given by $\mathcal{R} \triangleq \mathrm{log_2}(1+\gamma)$.

\subsection{Proposed Deep Learning Method}
\begin{figure}
  \begin{center}
  \includegraphics[width=82mm]{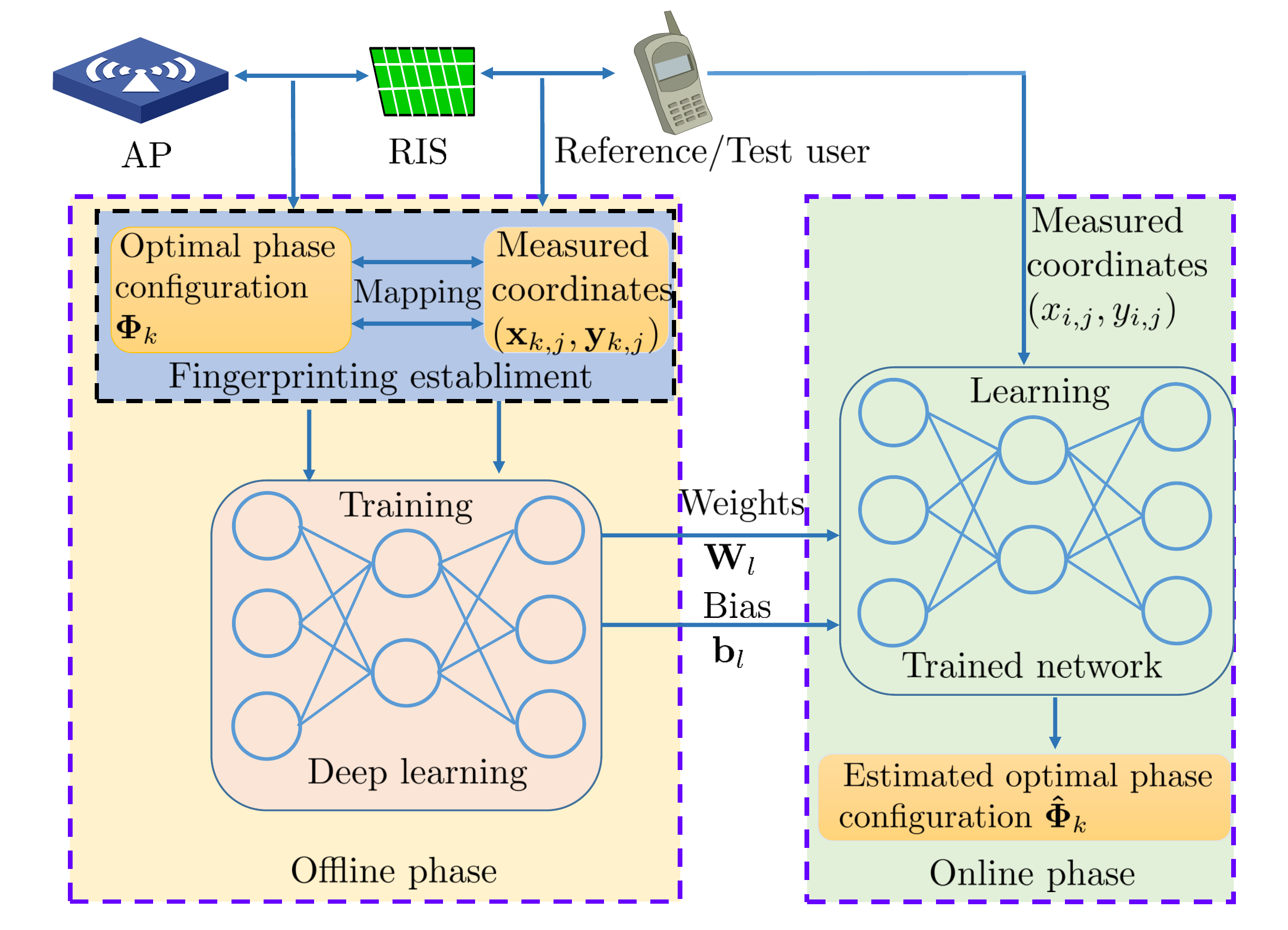}
	  \caption{The offline (training) and online phases of the proposed deep learning method for RIS-based indoor signal focusing improvement.}
  \label{fig:process} \vspace{-8mm}
  \end{center}
\end{figure}

Our goal in this paper is to design the optimal RIS configuration in terms of received signal strength maximization at any intended user's indoor location. Currently, state-of-the-art approaches on RIS design (e.g., \cite{LIS_globecom2018,LIS_twc2018}) assume ideal acquisition of the involved channels $\mathbf{h}_{2}$, $\mathbf{H}_{1}$, and $\mathbf{h}$, and adopt tools from optimization theory. However, at every channel coherence time, the latter approaches require complex operations for RIS configuration which are time and power consuming. To confront this issue, we propose a deep learning method for efficient online (re)configuration of intelligent surfaces. Our method capitalizes on the fact that the optimal phase configuration of the RIS elements for an intended indoor user depends on its position. The objective of the proposed DNN is to compute the mapping between the measured coordinate information of the targeted user and the optimal phase matrix $\mathbf{\Phi}$ that maximizes this user's received signal strength. More specifically, the designed DNN-based signal focusing system performs as illustrated in Fig$.$~\ref{fig:process}, with its operation composed of the following two distinct phases.
\begin{enumerate}
\item \textit{Offline (Training) Phase}: During this phase, a fingerprinting database is created that includes the measured coordinate information at various reference indoor user positions and their corresponding optimal phase configurations. Suppose $K$ measured coordinates, each denoted as $(x_k,y_k)$ with $k=1,2,\ldots,K$ and $\mathbf{\Phi}_k$ representing its corresponding optimal phase configuration. We assume that the coordinates at each RP location are computed via highly accurate indoor positioning techniques (e.g., \cite{Positioning}) and the optimal phase matrices are obtained via exhaustive search, or relevant sophisticated techniques (e.g, \cite{LIS_globecom2018}). After the creation of the fingerprinting database, its elements are used to train a DNN aiming at unveiling the relationship between the RPs' locations and the optimal RIS phase matrices. We assume that the fingerprinting database and DNN are placed on the RIS side; this requires dedicated storage and computing units to be attached to RIS via adequate interfacing. We further assume that RIS possesses a dedicated receiver for collecting information from RPs and the intended user.

\item \textit{Online Phase}: During the online phase of the proposed DNN-based method, the intended user first estimates its position's coordinates (e.g., via \cite{Positioning}) which are then communicated to RIS. Upon reception of this information, RIS feeds it to the designed DNN that outputs the required phase matrix for this user location. RIS adopts the computed phase configuration for assisting AP's transmission, until it receives new position coordinates from the intended user.
\end{enumerate}


\section{DNN-Based RIS Configuration}

\subsection{Deep Learning Basics}
Suppose that $\Theta\triangleq \{\bm{\theta}_{1},\bm{\theta}_{2},\ldots,\bm{\theta}_{L}\}$ includes $L$ sets of parameters. A feedforward DNN (or multi-layer perceptron) with $L$ layers describes a mapping $F(\mathbf{r}_0,\bm{\theta}): \mathbb{R}^{N_0\times1}\mapsto \mathbb{R}^{N_L\times1}$ of the input vector $\mathbf{r}_0 \in \mathbb{R}^{N_0\times1} $ to an output vector in $\mathbb{R}^{N_L\times1}$ through the following $L$ iterative processing steps:
\begin{equation}\label{model_DL1}
  \mathbf{r}_{\ell}\triangleq f_{\ell}( \mathbf{r}_{\ell-1};\bm{\theta}_{\ell}),\,\,\ell=1,2,\ldots,L,
\end{equation}
where $f_{\ell}( \mathbf{r}_{\ell-1};\bm{\theta}_{\ell}): \mathbb{R}^{N_0\times1}\mapsto \mathbb{R}^{N_L\times1}$ represents the mapping carried out by the $\ell$-th DNN layer. This mapping depends on the output vector $\mathbf{r}_{\ell-1}$ from the previous $(\ell-1)$-th layer and on a set of parameters $\bm{\theta}_{\ell}$. In general, the mapping $f_{\ell}(\cdot;\cdot)$ can be stochastic, i$.$e$.$, it can be a function of random variables. The $\ell$-th DNN layer is called dense or fully-connected if all neurons in this layer are connected to all neurons in the following layer. In this case, $f_{\ell}( \mathbf{r}_{\ell-1};\bm{\theta}_{\ell})$ with $\bm{\theta}_{\ell}\triangleq\{\mathbf{W}_{\ell},\mathbf{b}_{\ell}\}$ has the form:
\begin{equation}\label{model_DL2}
f_{\ell}( \mathbf{r}_{\ell-1};\bm{\theta}_{\ell})=\sigma(\mathbf{W}_{\ell}\mathbf{r}_{\ell-1}+\mathbf{b}_{\ell}),
\end{equation}
where $\mathbf{W}_{\ell} \in \mathbb{R}^{N_{\ell}\times (N_{\ell}-1)}$ denotes the neurons' weights at this layer, $\mathbf{b}_{\ell} \in \mathbb{R}^{N_{\ell}\times1}$ stands for the bias vector, and $\sigma(\cdot)$ represents an activation function.

\subsection{Fingerprinting Database Design}\label{sec:CSI_establishment}
For the creation of the fingerprinting database, we perform $J$ distinct estimates for each of the $K$ positions of the RPs. By gathering the $j$-th estimate (with $j=1,2,\ldots,J$) at all RP's position, we formulate the $j$-th set of the database:
\begin{equation}\label{Fingerprinting01}
  \Omega_j \triangleq \{(\mathbf{p}_{1,j}, \mathbf{\Phi}_{1,j}),(\mathbf{p}_{2,j}, \mathbf{\Phi}_{2,j}),...,(\mathbf{p}_{K,j}, \mathbf{\Phi}_{K,j})\},
\end{equation}
where $\mathbf{p}_{k,j}\triangleq[x_{k,j}\,y_{k,j}]$ contains the $j$-th estimate of the coordinates of the $k$-th RP position and $\mathbf{\Phi}_{k,j}\in\mathbb{C}^{N\times N}$ represents its corresponding optimal RIS phase configuration.

\subsection{Proposed DNN Design}\label{sec:proposed_DNN}
\begin{figure}
  \begin{center}
  \includegraphics[width=85mm]{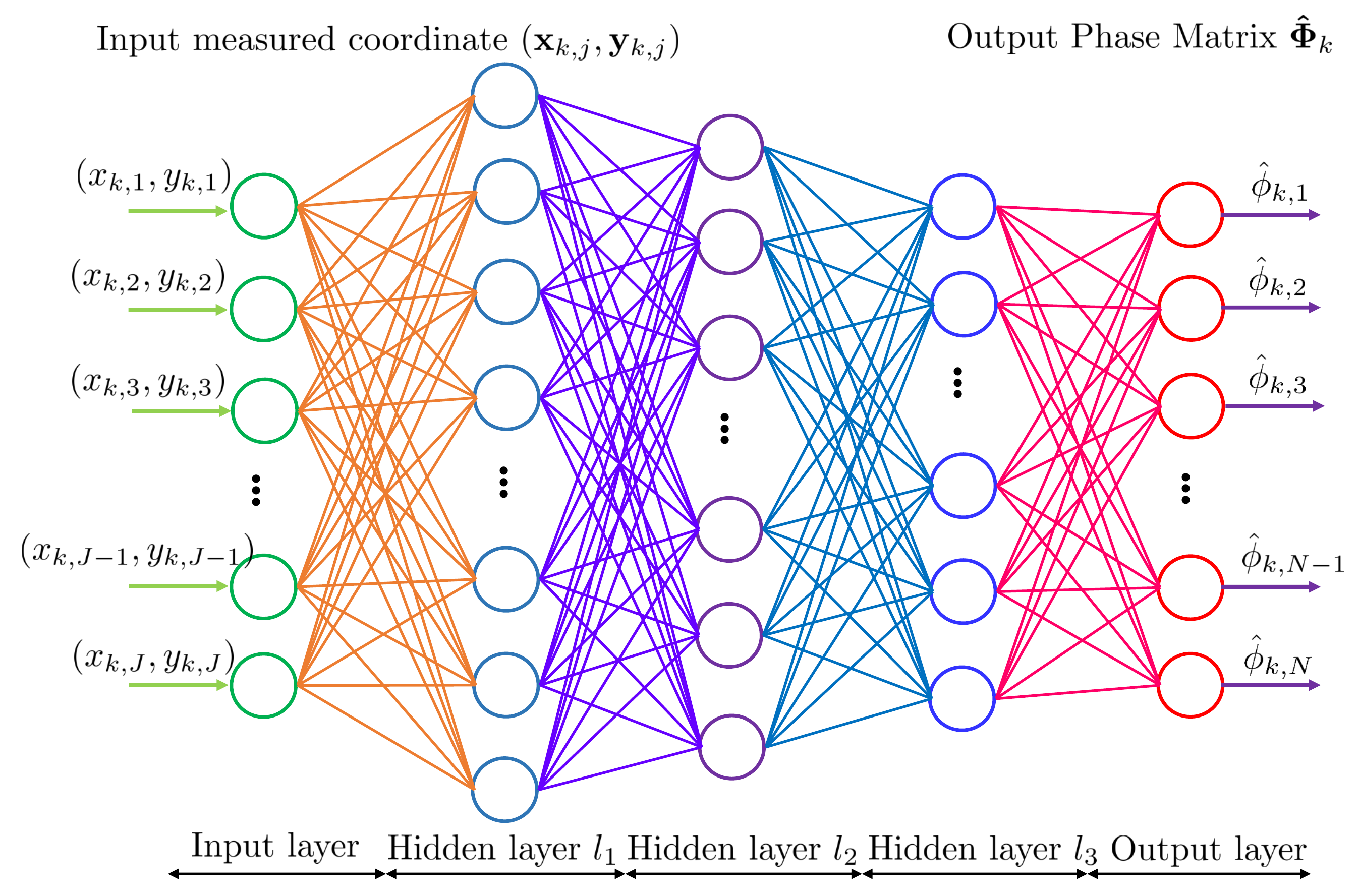}  \vspace{-2mm}
  \caption{The proposed five-layer DNN for the RIS phase matrix configuration. The DNN is fed with the estimation of the user position coordinates.}
  \label{fig:DNN} \vspace{-7mm}
  \end{center}
\end{figure}

According to the well-known \textit{universal approximation theorem} \cite{DNN_universal}, a feed-forward neural network with a single hidden layer processed by a multi-layer perception mechanism can approximate continuous functions on compact subsets of $\mathbb{R}^{N\times1}$. As a standard multi-layer processor, a DNN is capable of approximating any continuous function to any desired degree of accuracy. Our proposed DNN architecture for the RIS configurations yielding improved indoor signal focusing is illustrated in Fig$.$~\ref{fig:DNN}. It consists of five layers; namely, the input layer, three hidden layers, and the output layer. In particular, there exist multiple neurons in each hidden layer, and each layer's output is a nonlinear function of a weighted sum of the values of the neurons at the input of this layer. As shown in Fig$.$~\ref{fig:DNN}, the proposed DNN's input layer consists of $J$ neurons with values the estimated user position coordinates, which are propagated to the first hidden layer. The role of all three hidden layers is to capture the relationship between the estimation user position information and the optimal RIS phase configuration. Finally, we exploit the output layer to provide the estimation $\mathbf{\hat{\Phi}}$ for the optimal phase matrix by minimizing the difference between itself and the optimal $\mathbf{\Phi}$ (computed via exhaustive search or more efiicient optimization \cite{LIS_globecom2018}). In all layers, we have used $\mathrm{tanh}(\cdot)$ as our DNN's activation function $\sigma(\cdot)$, since it works well with approximations of nonlinear functions and provides full mapping in $[-1,1]$. In addition, $\mathrm{tanh}(\cdot)$ leads to stronger gradients compared with the $\mathrm{sigmoid}(\cdot)$ activation function.

\subsection{Proposed DNN Training Process}\label{sec:training}
In the beginning of the training phase, the $KJ$ desired inputs $(x_{k,j},y_{k,j})$'s and outputs $\mathbf{\Phi}_{k,j}$'s ($j=1,2,\ldots,J$ and $k=1,2,\ldots,K$) of the DNN are gathered via dedicated position estimation and optimization. Using this database information, the weights and bias vector of the DNN's five layers are trained according to a desired objective. We denote by $\mathbf{W}_{\ell}$ and $\mathbf{b}_{\ell}$ the weight matrix and bias, respectively, of the $\ell$-th layer with $\ell=1,2,\ldots,5$. Clearly, each $\mathbf{\hat{\Phi}}_{k,j}$ from the $KJ$ actual DNN outputs is a function of $(x_{k,j},y_{k,j})$'s and $\Theta\triangleq \{\bm{\theta}_{1},\bm{\theta}_{2},\ldots,\bm{\theta}_{5}\}$ with $\bm{\theta}_{\ell}\triangleq\{\mathbf{W}_{\ell},\mathbf{b}_{\ell}\}$. We have used the Minimum Mean Squared Error (MMSE) criterion among $\mathbf{\hat{\Phi}}_{k,j}$'s of the DNN and the respective desired outputs $\mathbf{\Phi}_{k,j}$'s:
\begin{align}\label{model_DL3}
    \mathcal{L}(\Theta) = \sum_{k=1}^{K}\sum_{j=1}^{J}\left\|\mathbf{\Phi}_{k,j}-\mathbf{\hat{\Phi}}_{k,j}(\Theta)\right\|_2^2,
\end{align}
where notation $\mathbf{\hat{\Phi}}_{k,j}(\Theta)$ indicates the dependence of a DNN's output to its layer's parameters.

To obtain $\Theta$ minimizing \eqref{model_DL3} using small batches of training data and with reasonable computational complexity, algorithms based on the stochastic gradient descent are usually employed. According to such algorithms, the DNN's parameters at its $\ell$-th layer are designed as
\begin{align}\label{training12}
\bm{\theta}_\ell = \bm{\theta}_{\ell-1} -\eta \nabla \mathcal{L}(\bm{\theta}_{\ell-1}),
\end{align}
where $\eta$ denotes the algorithm's learning rate and $\nabla \mathcal{L}(\mathbf{\cdot})$ is the gradient of the batch's loss function. For the design of the proposed DNN, we have used the Adagrad algorithm \cite{adam} that adopts the backpropagation method for efficient gradient computation.

\section{Numerical Results}
\begin{figure}
  \begin{center}
  \includegraphics[width=85mm]{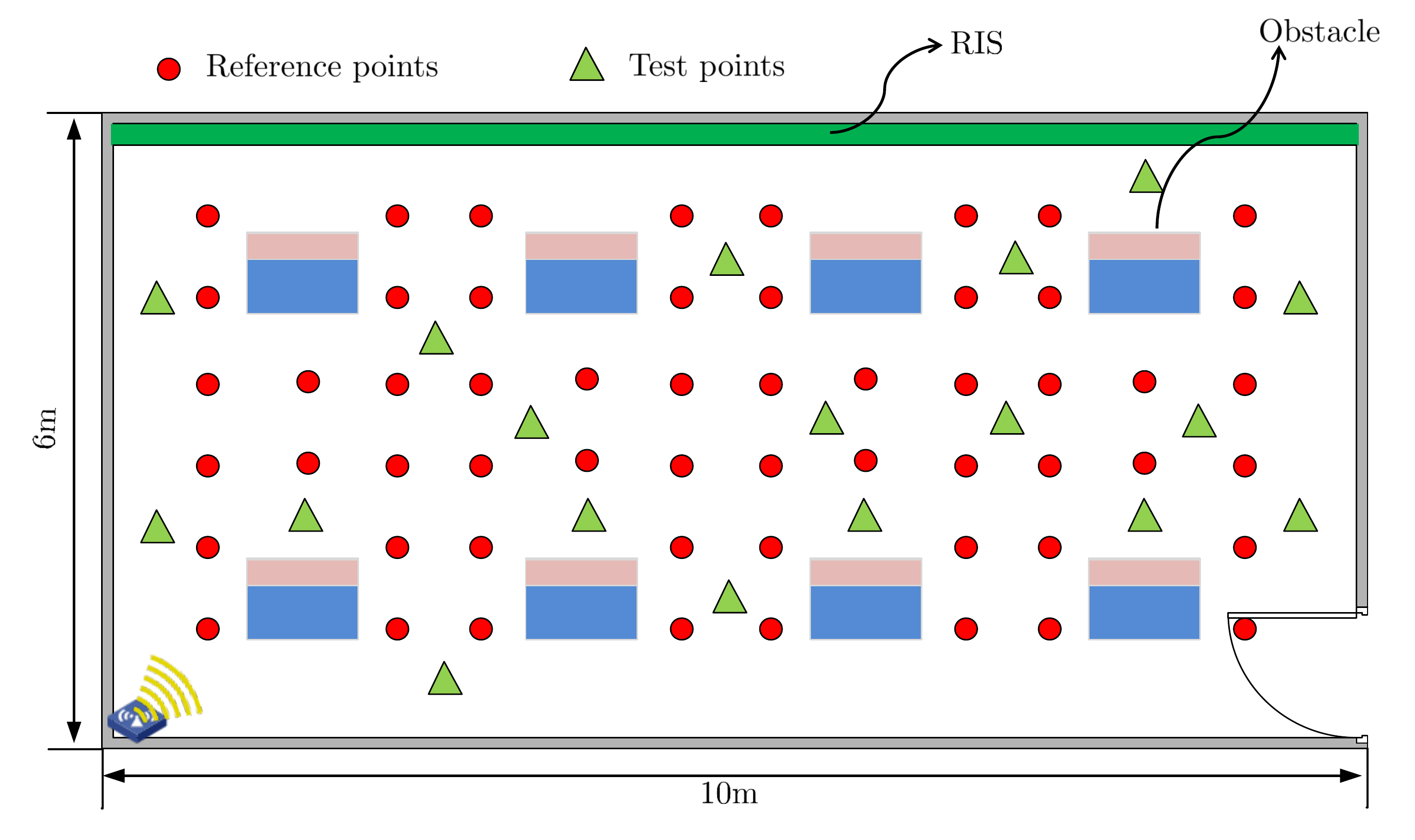}  \vspace{-6mm}
  \caption{Layout of the reference and test points as well as the RIS and obstacles' deployment for the considered indoor communication environment.}
  \label{fig:deployment} \vspace{-6mm}
  \end{center}
\end{figure}

\subsection{Simulation Parameters}
We have simulated a 3D office environment similar to the one depicted in Fig$.$~\ref{fig:lisindoor} and having $6m$ of width, $10m$ of length, and height $3m$. A RIS of $N$ reflecting elements with infinite resolution is coated on the upper wall of the figure, whereas an AP equipped with $M=32$ antenna elements is placed on the bottom left corner of the room's floor. In Fig$.$~\ref{fig:deployment}, we sketch the RPs whose estimated coordinates were used for creating the fingerprinting database needed in the proposed DNN's offline training phase. This figure also illustrates the positions of the obstacles inside the office, and the test points on the room's floor where the intended user was placed during the online RIS configuration phase. We have simulated $240$ different RP positions (some of them are drawn with red solid circles in Fig$.$~\ref{fig:deployment}) that were uniformly scattered with $0.5m$ spacing inside the room's floor. For the test points we have considered $32$ different positions; some of them are also depicted in the figure with green solid triangles. The room also includes $8$ uniformly located obstacles, each with height $1m$, width $0.6m$, and length $1m$.

In the results that follow we have assumed timely estimation of the intended user position during the DNN's online phase. For the wireless channels we have considered independent Rayleigh small-scale fading, and the pathloss was modeled as $20.4\mathrm{log}_{10}(d/d_0)$ with $d_0=1m$ denoting the reference distance and $d$ being the transmitter-receiver separation distance in meters \cite{indoormodel}. The variance of the additive white Gaussian noise was set to $1$. We have assumed a carrier frequency of $2.6$GHz, and that AP transmits with power ranging from $-10$dB to $30$dB an orthogonal frequency division modulated symbol with $150$kHz of bandwidth per subcarrier. To focus our investigation onto RIS contribution, we have assumed that it is the only structure in the room offering signal reflections (no futher reflections of AP's signal from the floor, ceiling or obstacles, neither from the other three walls not coated with a RIS. We have used $200$ Monte Carlo runs for the average performance results presented in the following Figs$.$~\ref{fig:compar} and~\ref{fig:numbk}.

\subsection{Performance Evaluation}
\begin{figure}
  \begin{center}
  \includegraphics[width=78mm]{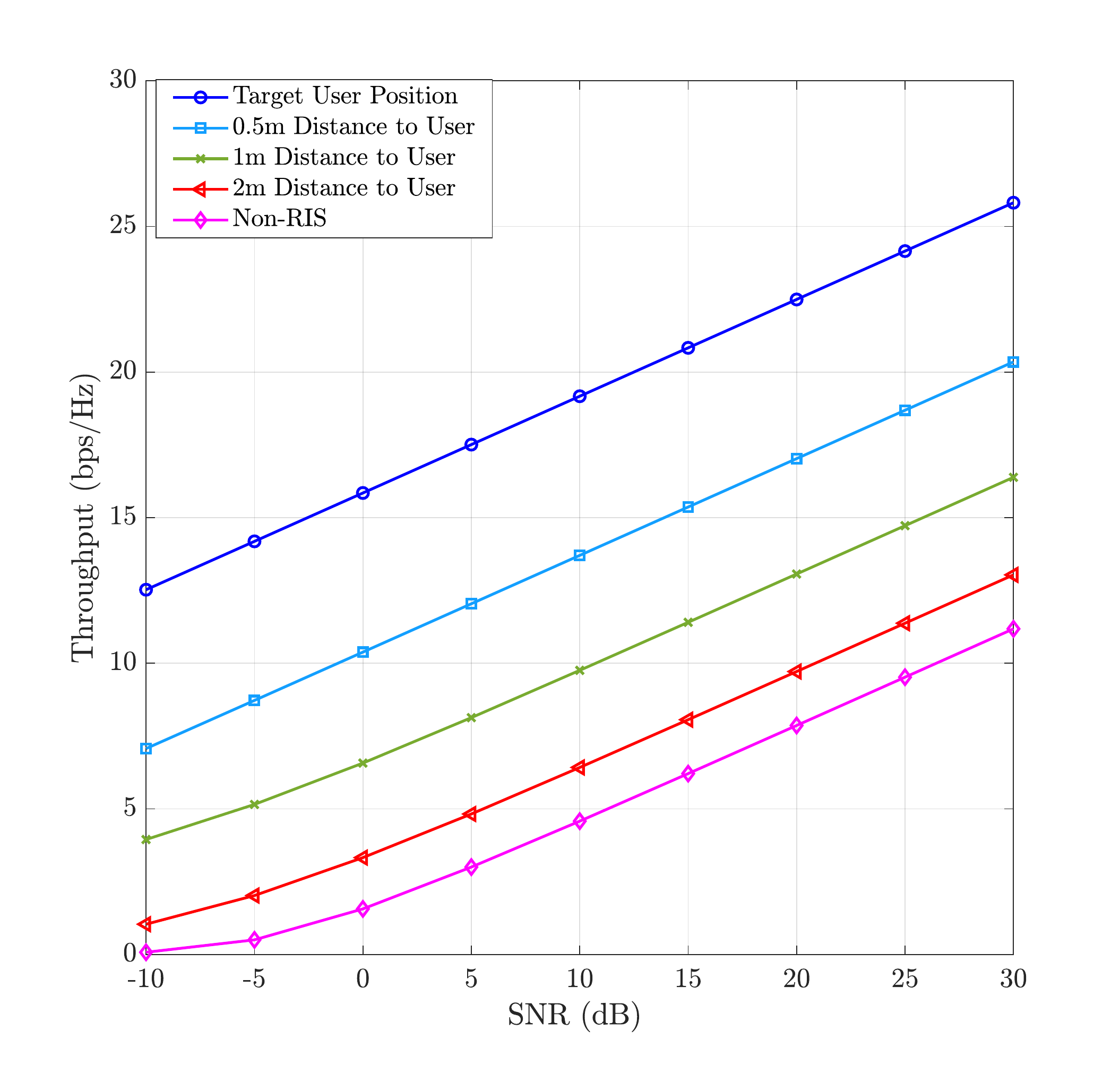}  \vspace{-4mm}
  \caption{Achievable throughput as a function of the transmit SNR $p/\sigma^2$ for the considered RIS-assisted indoor communication. The proposed DNN is deployed for the RIS configuration for the target user position. The throughputs at the distances $0.5m$, $1m$, and $2m$ from the target user are also sketched, as well as for the target user position for the non-RIS case.}
  \label{fig:compar} \vspace{-6mm}
  \end{center}
\end{figure}

\begin{figure}
  \begin{center}
  \includegraphics[width=83mm]{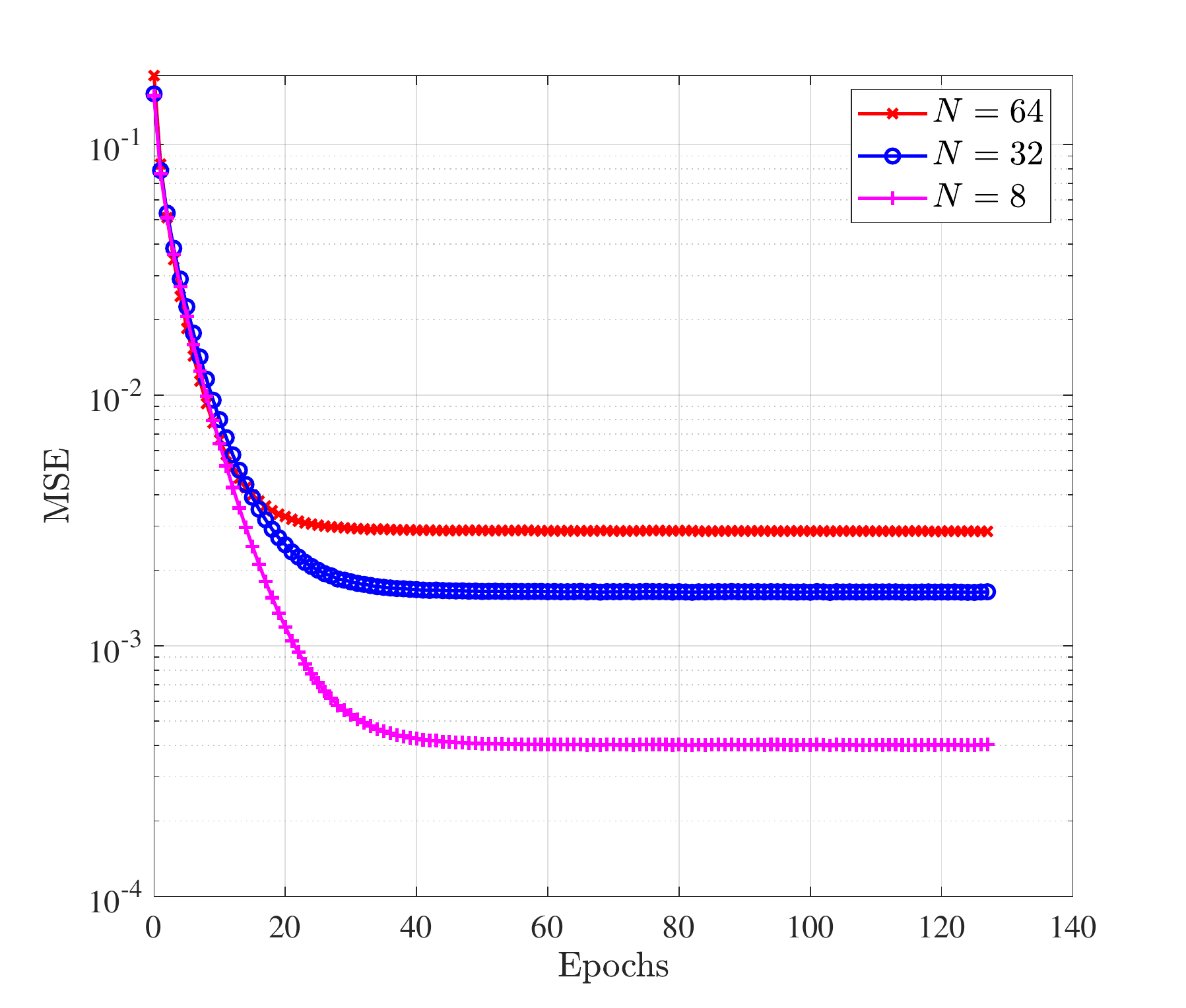}  \vspace{-4mm}
  \caption{MSE performance of the RIS configuration for an arbitrary test user position versus the number of the RIS reflecting elements $N$.}
  \label{fig:numbk} \vspace{-6mm}
  \end{center}
\end{figure}

We start with Fig.~\ref{fig:compar} that illustrates the signal focusing effect using RIS-assisted indoor communication. In particular, this figure depicts the achievable rate $\mathcal{R}$ versus SNR using the designed DNN with 50 epochs training for a given target user position.  The rate values at the distances $0.5m$, $1m$, and $2m$ further from this user location are also shown together with the case where AP's transmission to the intended user is not assisted by a RIS. As shown, RIS deployment substantially improves $\mathcal{R}$ yielding increased signal focusing at the target user position. The impact of the number RIS reflecting elements $N$ on the RIS configuration using the proposed DNN-based method is demonstrated in Fig$.$~\ref{fig:numbk}. In this figure, the MSE performance (as defined in \eqref{model_DL3}) between the output RIS phase matrix of the DNN and the optimal phase matrix for a given user position is sketched. It can be seen that the proposed method achieves good performance even for few (around $20$) training epochs for all tested $N$ values. It is also shown that the $N=8$ case yields the minimum MSE requiring, however, a few more epochs for convergence compared to the other $N$ values. As intuitively expected, the smaller the $N$ is, the less phase values need to be estimated, hence, the smaller the estimation error becomes.

\section{Conclusion}
In this paper, we presented a DNN-based method for online wireless configuration of RISs in indoor communication environments. Leveraging a properly designed fingerprinting database, the proposed DNN is trained during an offline phase to estimate the mapping between a user's position and the configuration of the RIS's unit cells that maximizes this user's received signal strength. The designed DNN is fed during the online phase with the estimation of the coordinate information at a user location to output the desired RIS configuration. Our simulation results on a 3D office environment showcased that deep learning configured intelligent surfaces can effectively focus signal transmissions to intended user locations yielding good MSE performance for the RIS phase matrix.

\bibliographystyle{IEEEbib}
\bibliography{reference}

\end{document}